\documentclass[12pt]{iopart}

\usepackage{graphics}
\begin{document}

\title[Observation of drift and diffusion processes in Ti/TiO$_x$/Ti memristive devices]{Observation of drift and diffusion processes\\ in Ti/TiO$_x$/Ti memristive devices 
				 prepared\\ by tip-induced oxidation}
\author{I. Batko  and M. Batkova} %\inst{1}% etc
\address{Institute of Experimental  Physics,
 Slovak   Academy  of Sciences, Watsonova 47,
 040~01~Ko\v {s}ice, Slovakia}%

\begin{abstract}

We demonstrate that memristive devices can be fabricated by 
		tip-induced oxidation of thin metallic films using atomic force microscope.
Electrical measurements of such prepared Ti/TiO$_x$/Ti test  structures 
		confirmed their  memristive behavior and inferred diffusion of oxygen vacancies in the TiO$_x$ barrier.
Consequent Kelvin probe force microscopy studies provided evidence for the diffusion, 
		as well as for expected oxygen vacancy drift.
Time evolution of the space distribution of the vacancies due to the diffusion process 
		revealed 	minute-scale (at least) retention times of the devices.
The work presents 
		technology alternative for fabrication of memristive nanodevices 	in geometry favouring
		advantageous scanning probe microscopy studies of their in-barrier processes, 
		as well as widely utilizable approach to search for novel oxide materials for 
	 	perspective memristive applications.   	

\end{abstract}

\maketitle
Four decades ago Chua provided strong arguments \cite{Chua1971} 
		that besides the resistor, inductor, and capacitor, 
		there should exist one more fundamental passive circuit element, so-called 
	 	memristor (shortcut for memory resistor). 
A characteristic property of this two terminal device is its resistance, 
		which depends on history of the  current  passing through it
		resulting in hysteretic current-voltage ($I$-$V$) curves.
Although over time many observations of hysteretic $I$-$V$ curves were reported, 
	connection of this phenomenon with memristive behavior 
	was done only recently by Strukov and co-workers \cite{Strukov2008Nat}, 
	who shoved, that memristance arises naturally in nanoscale systems 
	in which solid-state electronic and ionic transport 
	are coupled under an external bias voltage \cite{Strukov2008Nat}. 
As typical representatives of memristive devices can be considered  
		metal/oxide/metal two-terminal nanodevices, and
   Pt/TiO$_2$/Pt nanodevices (exhibiting bipolar resistive switching)
   seem to be a prototypical (nano)structure exhibiting memristive behavior.
Based on experimental studies of TiO$_2$ junction devices with platinum electrodes
	it was established that electrical conduction in metal/oxide/metal thin-film devices
	is controlled by a spatially heterogeneous metal/oxide electronic barrier
	\cite{YangNatureNano2008},
and memristive electrical switching proceeds by means of the drift of positively charged
	oxygen vacancies acting as native dopants to form (turn ON) or disperse (turn OFF)
	locally conductive switches through the electronic barrier
  \cite{YangNatureNano2008}.

Although metal/oxide/metal type devices exhibiting
		memristive/resistive switching are intensively investigated for very long time
		(especially because of their potential use
		in non-volatile memory applications), 	
		and significant progress in the field was done after ''rediscovery'' of the memristor
\cite{Strukov2008Nat,YangNatureNano2008,Pershin_Adv_Phys_2011,Kwon2010,Strachan_2011,Xia_Nanotechnology_2011,Robinett_Nanotechnology_2011,M-Ribeiro_Nanotechnology_2011,Waser2007,Yang_AdvMat2009,Strachan_AdvMat2010}, 
		there are still many open questions 
		related to local physical and chemical properties in the oxide barrier region \cite{Szot2011}.
Many of these peculiarities arising at nanoscale could be very effectively
		studied by various advantageous scanning probe microscopy (SPM) techniques. 
For instance, Kelvin probe force microscopy (KPFM), 
		which measures contact potential between the sample and a (metal coated) tip of atomic force 			microscope (AFM) \cite{Nonemacher_APL_1991}, and 
		is obviously used 	for direct detection 	of the work-function changes in metals or doping-level 			changes in semiconductors	\cite{Nonemacher_APL_1991,Meliz_2011}, 
	could provide valuable information about space distribution of oxygen vacancies
	in the active regions of the nanodevices.
Unfortunately, SPM studies of oxide-barrier processes in this kind of nanodevices   
		is a non-trivial task,
 		such as they are as a rule prepared in vertical, stacked geometry 
 		with oxide layer between a top and a bottom electrode 						\cite{YangNatureNano2008,Xia_Nanotechnology_2011,Yang_Nanotechnology_2009};
		thus removing of the ''top'' electrode
		is necessary to enable {\em direct} observation of the oxide barrier by SPM \cite{Yang_Nanotechnology_2009}.     
Stacked geometry of the memristive devices has at least one more negative aspect;
		it requires specific, often very expensive equipment
	that might be a limitation for widespread use in 
	searching for new materials for perspective memristive applications. 
Therefore,  
		new approaches enabling simple preparation of memristive devices  
		in geometry suitable for use of SPM-based techniques to study the  processes 
		in the oxide barrier are highly desired for faster progress in the field.

Interesting technology alternative for definition of oxide patterns with resolution in the 	nanoscale is 		local anodic oxidation (LAO) by use of AFM,
		which is well-established method for tip-induced oxidation of
		semiconducting \cite{Snow_APL_1993} and metallic \cite{Irmer_APL_1997} surfaces.
The method was succesfully used to fabricate many nanoelectronic and nanophotonic devices,
	e.g. metal/insulator/metal (MIM) type nanodevices, single-electron transistor, 
		or a photoconductive switch \cite{Matsumoto_1997}.
Important advantage of this method is 
		lateral planar geometry of devices fabricated this way,
		which enables realization of studies of the oxidized regions
 		by desired SPM techniques. 

The aim of this work is to demonstrate that LAO can be effectively used also for 
	fabrication of metal/oxide/metal (MIM) type devices exhibiting memristive behavior. 
In the presented case study Ti/TiO$_x$/Ti structures were prepared 
 	such as TiO$_2$ (an important fraction of TiO$_x$ formed in the LAO process 					    		\cite{Matsumoto_1997}), is a prototypical memristive material \cite{Szot2011}.
Consequently, KPFM was used to perform direct experimental observations of
 		  oxygen vacancy movement in the TiO$_x$ region.     
%%%%%%%%%%%%%%%%%%%%%%%%%%%%%%%%%%%%%%%%%%%%%%%%%%% 	
 	%\section{Experimental details}
 	
For purposes of this work, MIM type devices on the base of Ti thin films were prepared 
	as follows. 
First, microbridges (6-8 nm thick,  30-80 ~$\mu$m wide, 100-300~$\mu$m long) were 
  deposited by DC magnetron sputtering on glass substrates kept at ambient temperature.
The sputtering was done from polycrystalline Ti target
	at Ar-pressure of ~3 mTorr and rate of 0.04 nm/s.
A shadow mask was used to define the bridges. 
Oxide barrier across the bridge was fabricated by LAO 
  using AFM 
	equipped with commercial nanolithography software. 
The oxidation process was performed in contact mode at ambient conditions 
	with relative humidity between 55 and 60 \%. 
To ensure oxidation up to the substrate, 
 the lines across the bridge were overwritten several times  
 like described in Ref. 17, 20. %\cite{Irmer_APL_1997, Soltys_APP_2003}. 
The width of the formed TiO$_x$ lines varied between 400 and 1200~nm. 
Resistance of the films/microbridges before the oxidation was of order of tens of k$\Omega$;
  during the oxidation process, while forming Ti/TiO$_x$/Ti structure, 
	the resistance increased typically more than three orders of magnitude.
KPFM studies of the Ti/TiO$_x$/Ti structures were done in passive state 
	(with both Ti electrodes grounded), 
	as well as in active state (with bias voltage applied to one of the electrodes 
	while the other one was grounded). 
For electrical characterization of the devices 
	bias voltage, 	
	$V$, was applied to the prepared structure and the current, $I$, 
	flowing through it was measured. 

First electric characterization of the Ti/TiO$_x$/Ti devices was done 
	by measuring the $I$-$V$ curves at both increasing and decreasing $V$,
	at applied triangular excitation voltage signal of a period not longer
	than a few tens of seconds. 
Recorded curves (see inset in Fig. 1)	show almost symmetric characteristics 
	 strongly resembling tunneling between two identical metals. 
%
%
%%%%%%%%%%%%%%%%%%%%%%%%%
	\begin{figure}
				\begin{center}
											\resizebox{1\columnwidth}{!}{%
  			\includegraphics{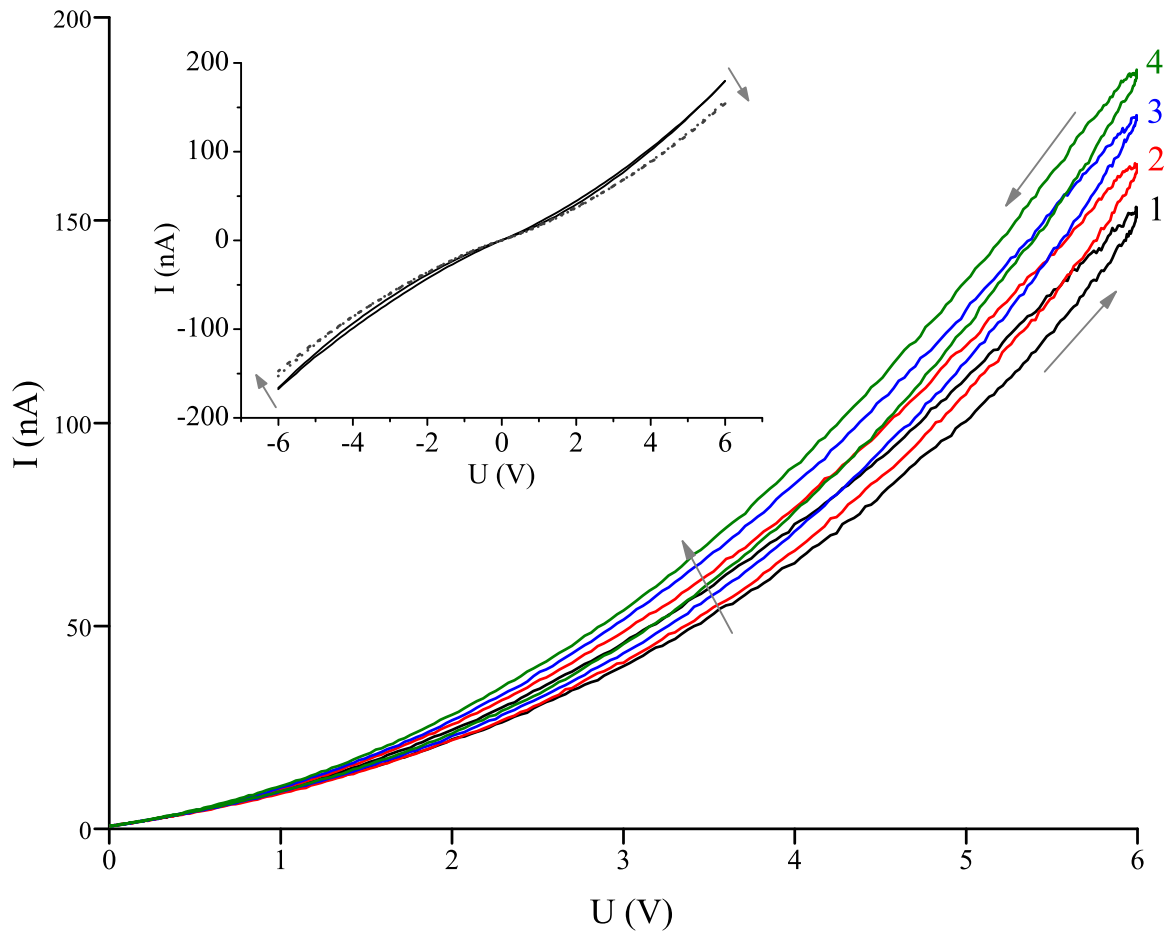}
            }
        \end{center}
\caption{ 
 $I$-$V$ hysteresis curves measured at alternating sinusoidal voltage with offset (see text) for four consecutive
	periods of $T=250$~s, showing continuous increase of conductance and overlapping of individual loops. 
Inset shows $I$-$V$ curves of the same device excited by (bipolar and symmetric) triangular waveform
	(of period 40~s).
	 }		\label{fig1}
			\end{figure}
%%%%%%%%%%%%%%%%%%%%%%%%%
%
%
In addition, the curves reveal very small hysteresis, 
		and a small decrease of the slopes, as shown in the inset of Fig.~1.
These  small, but well detectable changes reveal a change of the device conductance state
  due to the applied bias and are interpreted as a sign  of memristive behavior.
Considering a physical model of the memristor \cite{Pershin_Adv_Phys_2011,Strukov2008Nat} 
	a small hysteresis 
	can be adequately interpreted as a consequence of small change 
	of oxygen vacancy distribution in the TiO$_x$ region
	during one period of the applied voltage,
	what indicates relatively ''fast'' ramping 
	of the bias voltage \cite{Pershin_Adv_Phys_2011,Strukov2008Nat}. 
Exciting the device by positive voltage with sinusoidal modulation 
	and DC offset equal to the amplitude $A$ of the sinusoidal waveform 
	[$V(t) = A(1 + \sin(2\pi t/T)$,
	where $t$ is time and  
	$T$ is the period (now in the range of hundreds of seconds)], 
	 a sequence of $I$-$V$ loops clearly 
	resembling behavior expected for the memristor \cite{Strukov2008Nat} 
	was observed, 
 as can be seen in Fig.\ref{fig1}.
$I$-$V$ curves from four consequtive positive voltage sweeps
	show a continuous increase in conductance, 
	but also an overlap of the loops was observed. 
This is very similar observation like recently reported for Pd/WO$_3$/W devices \cite{Chang_ApplPhysA_2011},
	  interpreted as a consequence of 
		oxygen vacancy diffusion (in addition to their drift 
		under electric field) \cite{Chang_ApplPhysA_2011}.
The relevance of such interpretation was adequately supported by 
		simulations \cite{Chang_ApplPhysA_2011} using 
	the SPICE model of the memristor 
		\cite{Biolek_2009} 
	 incorporating both drift and diffusion into the model \cite{Chang_ApplPhysA_2011}.  
To interpret the observed overlap of $I$-$V$ loops in the Ti/TiO$_x$/Ti devices, 
		we adopt qualitatively the	same interpretation 
		(i.e. we consider that in addition to the drift of oxygen vacancies
 	 also their diffusion is substantional), 
		what we support by KPFM studies as described below.

%%%%%%%%%%%%%%%%%%%%%%%%%%%%%%%%
			\begin{figure}
						\begin{center}
							\resizebox{1.0\columnwidth}{!}{%
  			\includegraphics{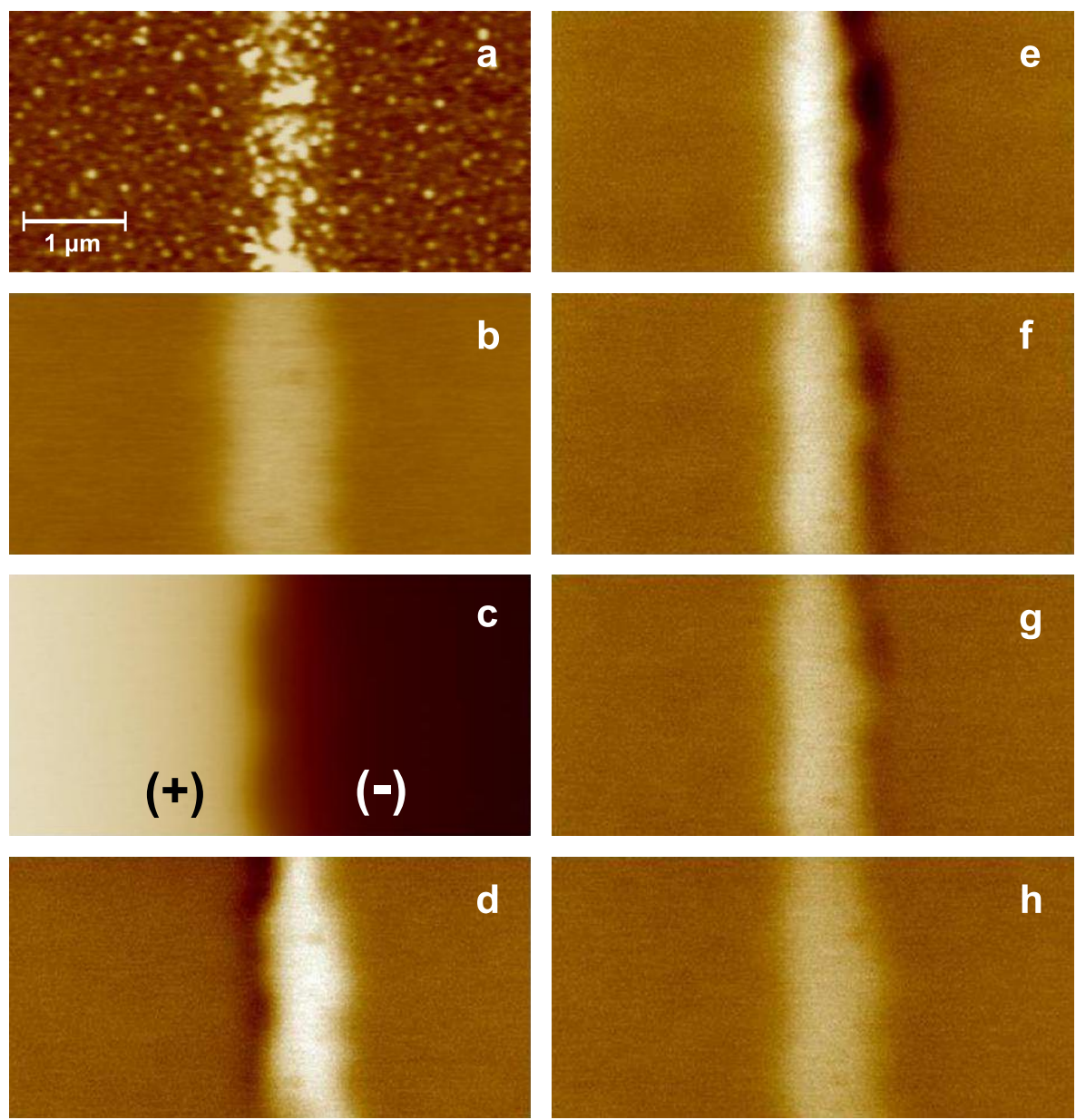}
            }
    	\vspace{-0.5cm}
 		\end{center}	
 			\caption{
Topography image (a) of Ti/TiO$_x$/Ti structure and corresponding
		surface potential images of the same structure before application
		of bias voltage (b), during applied bias voltage of +3~V (c),
		and immediately after removal the bias (d).
The surface potential of the same structure after voltage pulse 
		of the reverse polarity (-3~V) acquired immediately after the pulse (e), 
		and 8 (f), 17 (g), and 56 (h) minutes later. 
(Time duration of the acquisition of one image was approximately 4 minutes.)
The full colour-scale span for surface potential images is 220~mV, 
		except of c), where 3.5~V is used.
 }
\label{fig2}
\end{figure}
%%%%%%%%%%%%%%%%%%%%%%%%%%%%%%%%%%%%%
Initially, KPFM studies of Ti/TiO$_x$/Ti structure were done with both Ti electrodes grounded. 
Topography image of the region in the vicinity of the oxide barrier is shown in Fig. \ref{fig2}a, 
	while Fig. \ref {fig2}b shows the corresponding surface potential distribution measured by KPFM. 
Such as  the measurements were performed after the above mentioned electric studies 
	and no cleaning process was done to prevent change of the oxide barrier properties, 
	the topography image reveals quite degraded and contaminated surface.
In spite of this, 
	the surface potential image clearly reveals oxidized region of the width $\sim$1000~nm,
	 seen as a brighter region with  higher surface potential
		than that of Ti electrodes. 	 
Consequently, the bias voltage was applied between the Ti electrodes 
	and surface potential was measured for several values of the applied voltage 
	in the range of $0.5 - 3$~V. 
As demonstrated in Fig. \ref{fig2}c, at applied bias voltage of 3~V  
	(negative Ti electrode - cathode - is seen as a dark region)  
	 change of the surface potential is dominanting in the oxide barrier region.
%	 and well corresponds to the bias voltage applied to the device. 
% 
After the removal of the the bias voltage +3~V 
		the surface potential image (Fig.  \ref{fig2}d) 
		reveals formation of a wide  region of
	increased surface potential 
	in the oxide barrier on the side closer to the cathode 
	(bright region on the right side of the barrier),
		and a narrow region  with pronounced
	decrease of the potential on the opposite side of the oxide barrier 
	(dark region closer to the Ti electrode 
	that was biased positively - anode).	
The increase of the surface potential in the barrier region 
	we associate with increased concentration 
	of positively charged oxygen vacancies, 
	which drifted upon the applied voltage towards the cathode.
The decrease 	of the surface potential on the opposite side of the barrier we link to 
	 decreased concentration of positive oxygen vacancies and increased concentration 
	 of negative oxygen ions in the vicinity of the anode. 
(Note that concentration of negative oxygen ions in the vicinity of the anode
		can be also affected by formation of O$_2$ gas, 
		such as negative oxygen ions drifting to the anode can be discharged 
		to form O$_2$ gas there, as can be deduced from the results 
		of electroforming studies of Pt/TiO$_2$/Pt devices \cite{Yang_Nanotechnology_2009,Jeong_2008}.) 
	%%%%%%
Applying bias voltage of the opposite polarity (-3~V) for several minutes (not shown) 
		leads to inverse surface-potential picture (Fig. \ref{fig2}e), 
		thus confirming a voltage-driven drift of oxygen vacancies
		 during the application of bias voltage. 
As can be seen in figures \ref{fig2}e-h, the voltage-induced regions of
	higher/lower surface potential vanish with time,
	revealing equilibration process of the vacancy concentration in the barrier.
In accordance with the above discussed indications from the overlap
	of $I$-$V$ hysteresis loops we interpret  the observed equilibration process
	as a consequence of oxygen vacancy diffusion.
Nearly uniform surface potential, indicating almost uniform distribution 	
		of oxygen vacancies in the barrier
	  was observed again approximately an hour after the removal the bias voltage, 
	  as can be seen in Fig. \ref{fig2}h.

The above mentioned KPFM
		observations are considered as a direct evidence for the vacancy drift and diffusion
		in Ti/TiO$_x$/Ti memristive structures.
The corresponding time constant of the diffusion process 
	is estimated to be of order of tens of minutes. 
Such as the diffusion process in fact defines
	the time scale at which the device is capable to store an information, or when it loses its state \cite{Strukov_2009_ApplPhysA}, 
  the results of KPFM can also be interpreted  as indication for minute-scale (at least) retention times of the Ti/TiO$_x$/Ti memristive devices. 
Such retention times are applicable e.g. in bio-inspired circuits 
	 	 \cite{Strukov_2009_ApplPhysA}.

Here we would like to mention that in accordance to our knowledge
		the above presented results represent first 
		investigation of memristive device(s) fabricated by
		local anodic oxidation, 
		as well as first 
		direct experimental evidence for both vacancy drift and diffusion in the
		active part of a  memristive device.
	 %?? (by use of KPFM) ??.
Moreover, the results provide an experimental support 
		for the model proposed by other authors 
		\cite{Chang_ApplPhysA_2011}, who incorporated both drift and diffusion effects 
		into their SPACE simulation
		to explain the observed memristive behaviors of Pd/WO$_3$/W devices.    
The results also infer that KPFM can be used for {\em quantitative} 
			characterization of the diffusion processes in nanodevices 
			by scanning the time evolution of the surface potential 
			of the relevant part of the nanodevice after the removal 
			of the voltage applied to the device.
For studies of relatively slow diffusion processes
		 a repetitive scanning can be done for relatively larger area 
			just as done in this work. 
However, if more detailed information or highest ''sampling'' rates 
		of the scan region is required,  
		then scanning process can be reduced
			just to one scan line. 
The later approach should be preferred also in studies where the most detailed information
			about the diffusion dynamics in the selected surface cross-section of the active part of the
			device is required. 
		 
Important technological impact of the presented work is 
		that it shows a simple approach to search for new oxide materials
		with potential use in memristive applications (e.g. those for memristive/resistive switching ones). 
Such materials can be grown in the process of LAO, where very diverse 
	   materials can be used as precursors for the oxidation.
This approach is especially suitable for synthesis of complex oxides, 
		e.g.  by oxidation of multicomponent thin films. 
For instance, thin film precursors can be prepared by magnetron co-sputtering from several targets, 
		or may consist of several layers of different materials
		deposited/grown by any desired thin film technique.
Feasibility tests of oxide materials grown in the barrier of MIM structure
	  can be then done by means of electric measurements or/and by relevant SPM studies.

The described approach brings also interesting possibilities to studies 
		of electroforming, what is important technological step used
		to initialize resistive switching process in many oxides			
\cite{Yang_Nanotechnology_2009,Jeong_2008,Choi2005,Jeong2007,Gibbons1964,Hickmott1969,Szot2006,Oligschlaeger2006}.
Especially studies of electroforming combined with 
		advantageous SPM studies could shed some light on the processes in the oxide barrier 
		that are responsible for initiation of resistive switching. 
(Note that no electroforming was performed for Ti/TiO$_x$/Ti structures reported here,
		but preparation of such experiments is in progress.)

In summary, we managed to fabricate Ti/TiO$_x$/Ti devices
		exhibiting features of memristive behavior
		using the tip-induced oxidation,	
		thus providing geometry favourable for investigations
		of phenomena in the oxide barrier by use of SPM techniques.
Performed electric measurements revealed memristive behavior of the devices, 	
			and indicated importance of diffusion processes in the TiO$_x$ barrier. 
Kelvin probe force microscopy studies provided direct evidence for voltage-induced drift 
		and consequent 			diffusion of oxygen vacancies,  
		and revealed retention times of the devices at minute-scale, thus inferring that
	  KPFM can be used for {\em quantitative} 
			characterization of the diffusion processes in materials and nanodevices. 
The presented approach represents 
	  a favorable technology alternative for routine fabrication of memristive devices
	  and investigation of physical phenomena in the active part of the device 
	   by use of advantageous SPM techniques. 
Such as MIM type structures with many new oxide materials
		in the oxide barrier 
		can be routinely fabricated by tip-induced oxidation,
		we believe that presented approach will be widely
		used to seek for new oxide materials for perspective memristive applications.

%%%%%%%%%%%%%%%%%%%%%%%%%%%%%%%%%%%%%%%%%%%%%%%%%%%%%%%%%%%%%%%%%%%%%%%%
 \section*{Acknowledgments}
 This work was supported by 
the Slovak Scientific Agency VEGA under the contract No. 2-0133-09, 
and by the ERDF EU (European Union European regional
development fond) grants, under the contract No. ITMS 26220120005 and ITMS 26220120047.

\section*{References}
%\bibliographystyle{unsrt}  %{unsrt}

% ****** End of file apssamp.tex ******
%%%%%%%%%%%%%%%%%%%%%%%%%%%%%%%
%%%%%%%%%%%%%%%%%%%%%%%%%%%%%%%%
%
%\begin{thebibliography}{}
%%
%% and use \bibitem to create references.
%Author, \textit{Book title} (Publisher, place year) page numbers
%\end{thebibliography}

% \end{thebibliography}

%\endrefs


\begin{thebibliography}{50}
%\end{thebibliography}
% 
% %\bibliography{references-Memr}
% 

\bibitem{Chua1971}
Chua L~O  1971 \textit{IEE Trans. Circuit Theory} \textbf{18} 507

\bibitem{Strukov2008Nat}
 Strukov D~B et al 2008 \textit{Nature} (London) \textbf{453} 80

\bibitem{YangNatureNano2008}
Yang J~J et al 2008 \textit{Nat. Nanotechnol.} \textbf{3} 429

\bibitem{Pershin_Adv_Phys_2011}
Pershin Yu~V and  Di Ventra M 2011 \textit{Adv. Phys.} \textbf{60} 145
 
\bibitem{Kwon2010}
 Kwon D-H et al 2008 \textit{Nat. Nanotechnol.} \textbf{3} 429

\bibitem{Strachan_2011}
Strachan J~P et al 2011 \textit{Nanotechnology} \textbf{22} 254015

\bibitem{Xia_Nanotechnology_2011}
 Xia Q et al 2011  \textit{Nanotechnology} \textbf{22} 254026

\bibitem{Robinett_Nanotechnology_2011}
Robinett W. 2010  \textit{Nanotechnology} \textbf{21} 235203

\bibitem{M-Ribeiro_Nanotechnology_2011}
 Medeiros-Riberio G  2011 \textit{Nanotechnology} \textbf{22} 095702

\bibitem{Waser2007}
 Waser R and  Aono M 2007 \textit{Nat. Mater.} \textbf{6} 833

\bibitem{Yang_AdvMat2009}
 Yang J~J et al 2009  \textit{Adv. Mater.} \textbf{21} 3754 
 
\bibitem{Strachan_AdvMat2010}
 Strachan J~P et al  2010 \textit{Adv. Mater.} \textbf{22} 3573  

\bibitem{Szot2011}
 Szot K 2011  \textit{Nanotechnology} \textbf{22} 254001

\bibitem{Nonemacher_APL_1991}
 Nonnenmacher N et al 1991 \textit{Appl. Phys. Lett.} \textbf{58} 2921

\bibitem{Meliz_2011}
 Melitz W 2011 \textit{Surface Science Reports} \textbf{66} 1

\bibitem{Yang_Nanotechnology_2009}
Yang J~J et al 2009 \textit{Nanotechnology} \textbf{20} 215201
 
\bibitem{Irmer_APL_1997}
 Irmer B et al 1997 \textit{Appl. Phys. Lett.} \textbf{71} 1733

\bibitem{Snow_APL_1993}
 Snow E~S et al 1993 \textit{Appl. Phys. Lett.}  \textbf{63} 749
%
\bibitem{Matsumoto_1997}
 Matsumoto K 1997 \textit{Proc. IEEE} \textbf{85} 612

\bibitem{Soltys_APP_2003}
\v {S}olt\'ys J.  et al 2003 \textit{Acta Phys. Polonica A} \textbf{103} 553

\bibitem{Chang_ApplPhysA_2011}
 Chang T et al 2011 \textit{Appl. Phys. A}  \textbf{102} 857

\bibitem{Biolek_2009} 
 Biolek Z et al 2009 \textit{Radioengeneering} \textbf{18} 210

\bibitem{Strukov_2009_ApplPhysA}
 Strukov D~B and  Wiliams R~S 2009 \textit{Appl. Phys. A}  \textbf{94} 515

\bibitem{Jeong_2008}
 Jeong et al 2008 \textit{J. Appl. Phys.} \textbf{104} 123716

\bibitem{Choi2005}
 Choi B~J et al 2005 \textit{J. Appl. Phys.} \textbf{98} 033715

\bibitem{Jeong2007}
 Jeong D S et al 2007 \textit{Electrochem. Solid-State Lett.} \textbf{10} G51

\bibitem{Gibbons1964}
Gibbons J~F and  Beadle W~E 1964 \textit{Solid-State Electron.} \textbf{7} 785

\bibitem{Hickmott1969}
 Hickmott T~W \textit{J. Vac. Sci. Technol.} \textbf{6} 828

\bibitem{Szot2006}
 Szot K et al 2006  \textit{Nat. Mater.} \textbf{5} 312

\bibitem{Oligschlaeger2006}
Oligschlaeger R.  et al 2006 \textit{Appl. Phys. Lett.}  \textbf{88} 042901

 
\end{thebibliography}
\end{document}